
%
\input harvmac.tex

\def\grad{{\rm grad~}}
\def\Ht{{\hat H}}
\def\P {P}
\def\CO {{\cal O }}
\def\CJ {{\cal J }}
\def\CP {{\cal P }}
\def\CC {{\cal C }}

\def\p {\partial}

\def\inbar{\,\vrule height1.5ex width.4pt depth0pt}
\def\IB{\relax{\rm I\kern-.18em B}}
\def\IC{\relax\hbox{$\inbar\kern-.3em{\rm C}$}}
\def\IP{\relax{\rm I\kern-.18em P}}
\def\IR{\relax{\rm I\kern-.18em R}}

\def\inbar{\,\vrule height1.5ex width.4pt depth0pt}
\def\IB{\relax{\rm I\kern-.18em B}}
\def\IC{\relax\hbox{$\inbar\kern-.3em{\rm C}$}}
\def\ID{\relax{\rm I\kern-.18em D}}
\def\IE{\relax{\rm I\kern-.18em E}}
\def\IF{\relax{\rm I\kern-.18em F}}
\def\IG{\relax\hbox{$\inbar\kern-.3em{\rm G}$}}
\def\IH{\relax{\rm I\kern-.18em H}}
\def\II{\relax{\rm I\kern-.18em I}}
\def\IK{\relax{\rm I\kern-.18em K}}
\def\IL{\relax{\rm I\kern-.18em L}}
\def\IM{\relax{\rm I\kern-.18em M}}
\def\IN{\relax{\rm I\kern-.18em N}}
\def\IO{\relax\hbox{$\inbar\kern-.3em{\rm O}$}}
\def\IP{\relax{\rm I\kern-.18em P}}
\def\IQ{\relax\hbox{$\inbar\kern-.3em{\rm Q}$}}
\def\IR{\relax{\rm I\kern-.18em R}}
\font\cmss=cmss10 \font\cmsss=cmss10 at 7pt
\def\IZ{\relax\ifmmode\mathchoice
{\hbox{\cmss Z\kern-.4em Z}}{\hbox{\cmss Z\kern-.4em Z}}
{\lower.9pt\hbox{\cmsss Z\kern-.4em Z}}
{\lower1.2pt\hbox{\cmsss Z\kern-.4em Z}}\else{\cmss Z\kern-.4em Z}\fi}
\def\IGa{\relax\hbox{${\rm I}\kern-.18em\Gamma$}}
\def\IPi{\relax\hbox{${\rm I}\kern-.18em\Pi$}}
\def\ITh{\relax\hbox{$\inbar\kern-.3em\Theta$}}
\def\IOm{\relax\hbox{$\inbar\kern-3.00pt\Omega$}}

\magnification1200

{\nopagenumbers\abstractfont\hsize=\hstitle%
\rightline{\vbox{\baselineskip12pt\hbox{YCTP-P25-91}\hbox{RU-91-36}}}%
\vskip .3in\centerline{\titlefont%
\vbox{\centerline{Loop Equations and the Topological Phase}}}%
\vskip .2in\centerline{\titlefont%
\vbox{\centerline{of Multi-Cut Matrix Models}}}%
\abstractfont\vskip .3in\pageno=0}

\centerline{\v C. Crnkovi\'c$^1$}
\centerline{M. Douglas$^2$}
\centerline{G. Moore$^3$}
\medskip
\centerline{$^{2}$ Dept. of Physics and Astronomy}
\centerline{Rutgers University}
\centerline{Piscataway, NJ 08855-0849}
\medskip
\centerline{$^{1,3}$ Department of Physics}
\centerline{Yale University}
\centerline{New Haven, CT 06511-8167}
\noindent
\bigskip
\baselineskip 18pt
\noindent

We study the double scaling limit of mKdV type,
realized in the two-cut Hermitian matrix model.
Building on the work of Periwal and Shevitz and of Nappi,
we find an exact solution including all odd scaling operators,
in terms of a hierarchy of flows
of $2\times 2$ matrices.
We derive from it loop equations
which can be expressed as Virasoro constraints on the
partition function.
We discover a ``pure topological" phase of the theory in which
all correlation functions are determined by recursion relations.
We also examine macroscopic loop amplitudes, which suggest a
relation to 2D gravity coupled to dense polymers.

\footnote{}{$^1$ (crnkovic@yalphy.hepnet, or \ @yalehep.bitnet),
Address after Sept. 1991: CERN, Theory Division, CH-1211, Geneva 23,
Switzerland. On leave of absence from Institute ``{Ruder Bo\v skovi\'c}
,'' Zagreb, Yugoslavia.}
\footnote{}{$^2$ (mrd@ruhets.rutgers.edu)}
\footnote{}{$^3$ (moore@yalphy.hepnet, or \ @yalehep.bitnet)}

\Date{August 15, 1991}

\newsec{Introduction}

Recent studies of two-dimensional
Euclidean gravity have yielded a host of surprises.
These surprises have come from lattice definitions
\nref\Ambjorn{ J. Ambj{\o}rn, B. Durhuus and J. Fr\"ohlich,
``Diseases of triangulated random surface models, and possible cures,''
Nucl.Phys. {\bf B257 [FS14]} (1985) 433.}%
\nref\David{ F. David, ``Planar diagrams, two-dimensional lattice
gravity and surface models,''
Nucl.Phys. {\bf B257 [FS14]} (1985) 45.}%
\nref\Migdal{ V. A. Kazakov, I. K. Kostov and A. A. Migdal,
``Critical properties of randomly triangulated planar random
surfaces,'' Phys. Lett. {\bf 157B} (1985) 295.}%
\nref\BKaz{ E. Br\'ezin and V. Kazakov, ``Exactly solvable field
theories of closed strings,'' Phys. Lett. {\bf B236} (1990) 144.}%
\nref\DSh{ M. Douglas and S. Shenker, ``Strings in less than one
dimension,'' Nucl. Phys. {\bf B335} (1990) 635.}%
\nref\Gmigdal{ D. Gross and A. Migdal,``Nonperturbative
two dimensional quantum gravity,'' Phys. Rev. Lett. {\bf 64} (1990)}%
\refs{\Ambjorn{--}\Gmigdal},
continuum path-integral definitions
\nref\Poly{ A.M. Polyakov, ``Quantum geometry of bosonic strings,''
Phys. Lett. {\bf 103B} (1981) 207.}%
\nref\KPZ{ A.M. Polyakov, ``Quantum gravity in two dimensions,'' Mod.
Phys. Lett. {\bf A2} (1987) 893\semi
V. Knizhnik, A. Polyakov and A. Zamolodchikov,
``Fractal structure of 2d-quantum gravity,''
Mod. Phys. Lett. {\bf A3} (1988) 819.}%
\nref\DDK{ F. David, ``Conformal field theories coupled to 2-D gravity
in the conformal gauge,''
Mod. Phys. Lett. {\bf A3} (1988) 1651; J. Distler
and H. Kawai, ``Conformal field theory and 2-D quantum gravity or
Who's afraid of Joseph Liouville?,''
Nucl. Phys. {\bf B321} (1989) 509.}
\refs{\Poly{--}\DDK}, and topological field theory definitions
\nref\witten{ E. Witten, ``On the structure of the topological phase
of two dimensional gravity,'' Nucl. Phys. {\bf B340} (1990) 281.}%
\nref\distler{ J. Distler, ``2D quantum gravity, topological field
theory and multicritical matrix models,'' Nucl. Phys. {\bf B342}
(1990) 523.}%
\nref\WD{ R. Dijkgraaf and E. Witten, ``Mean field theory, topological
field theory, and multi-matrix models,''
Nucl. Phys. {\bf B342}(1990)486.}
\refs{\witten{--}\WD} of what is presumably the same theory.

Among the surprises from the lattice/matrix-model approach
has been the discovery of
theories which look very similar to the solved models of
two-dimensional gravity coupled to matter, but for which no
continuum or topological interpretation has been found.
For example, a
particularly natural family of matrix models can be derived from the
integral over a unitary matrix.
These have a double scaling limit whose exact solution is strikingly
similar to that of the lattice gravity models, with the KdV hierarchy
being replaced by the mKdV hierarchy
\nref\peri{V. Periwal and D. Shevitz, ``Unitary-Matrix Models
as Exactly Solvable String Theories,'' Phys. Rev. Lett.
{\bf 64} (1990) 1326.}%
\nref\Pcor{V. Periwal and D. Shevitz, ``Exactly solvable unitary
matrix models: multicritical potentials and correlations,'' Nucl. Phys.
{\bf B344} (1990) 731.}%
\refs{\peri,\Pcor}.
One is naturally led to ask whether these theories have equally
simple world-sheet interpretations.

The simplest way to attack this problem
would be to find an interpretation of the
perturbation expansion for the theory as a lattice gravity theory,
possibly with matter and appropriate weights in the sum over graphs,
and directly take the world-sheet continuum limit.
This can be difficult since the same continuum
theory can be obtained from many different matrix models.
The important point is not the type of matrix but rather the
structure of the saddle-point eigenvalue distribution near its
points of non-analyticity.
The mKdV scaling limits come from tuning a coupling to a critical
point where two end-points of the eigenvalue distribution collide.
This is generic for a unitary matrix, but is easy to arrange
for a Hermitian matrix as well, for example by taking a double-well
potential
\nref\flow{ M. Douglas, N. Seiberg, and S. Shenker, ``Flow and
instability in quantum gravity,'' Phys. Lett.
{\bf B244}(1990)381}%
\nref\CCGM{ \v C. Crnkovi\'c and G. Moore, ``Multicritical multi-cut
matrix models,'' Phys. Lett. {\bf
B257}(1991)322}%
\refs{\flow,\CCGM}.
Thus there are many possible graphical formalisms which
one might try to interpret.
Perturbation theory for the unitary matrix integral can be done
by changing variables $U=\exp iH$; the resulting graphs have vertices
of arbitrary order as well as fermionic lines coming from a
fermionic representation of the Jacobian
\ref\Neuberger{ H. Neuberger, ``Scaling regime at
the large $N$ phase transition of two dimensional pure gauge
theories,'' Nucl. Phys. {\bf B340} (1990) 703.}.
We will discuss this expansion further in the conclusions.
Alternatively one could sum over the expansions about the various
minima of the double-well problem.

It is not obvious how to interpret any of these lattice theories,
because we do not know how to take the continuum limit directly
on the world-sheet. The simplest lattice gravity
\refs{\Ambjorn{--}\Gmigdal}\
is exceptional since it has the correct degrees of freedom
and an everywhere positive measure. Nevertheless,
already with non-positive measures there are
subtleties in the continuum limit which can reproduce gravity
coupled to matter
\nref\Kaza{V. A. Kazakov, Mod. Phys. Lett. {\bf A4} (1989) 2125.}%
\nref\Stau{M. Staudacher, ``The Yang-Lee Edge Singularity on a
Dynamical Planar Random Surface,'' Nucl. Phys. {\bf B336}(1990)349.}%
\refs{\Kaza,\Stau}.
The measures in the present problem are too complicated for
heuristic arguments.
All this underscores the importance of developing a more precise
understanding of this continuum limit.

Since we cannot deduce the continuum theory directly from
the lattice we are forced to compare the structure
of the exact solution of the lattice theory
with the structure of known continuum theories.
With this motivation, we describe below
the exact solution of the lattice theory in some detail.
In section two we derive from
the lattice the complete string and flow equations for two-cut models,
including all even and odd perturbations in the potential.
In section three we show how these equations can be
rewritten as Virasoro-type constraints. This reformulation
suggests the existence of a topological phase of the theory,
discussed in section four. In section five we make some
brief remarks about macroscopic loops in these theories.
In the conclusions we attempt to draw some inferences
about what the continuum formulation of these models might
be.

\newsec{Multicritical Matrix Model Potentials and Physical Observables}

In this section we derive the string equation and flow formalism,
generalizing \peri, \Pcor, \CCGM\ and
\ref\nappi{C. Nappi, ``Painlev\'e II and Odd Polynomials,''
IASSNS-HEP-90/61}.
The analysis is rather technical but the results are summarized in the
string equation (2.23), recursions (2.8), observables (2.19)
and flows (2.11).

To derive the full continuum loop equations for unitary matrix models,
we will work with the two-cut Hermitian matrix model and study the
most general scaling potential. The general even multicritical
potential resulting in a two-cut eigenvalue distribution was obtained
in \CCGM:
\eqn\mceven{V'_m(\lambda) = k(m) \lambda^{2m+1}\biggl( 1-{1\over
\lambda^2}\biggr)^{1/2}\biggl|_+,\qquad m=1,2,\ldots,}
where the subscript $+$ means we keep the polynomial part in an
expansion about infinity and the constant
$k(m)=2^{2m+1}(m+1)!(m-1)!/(2m-1)!$.

The universality class of a potential is determined by the behaviour
of the saddle-point eigenvalue density $\rho(\lambda)$ around the
critical point. The eigenvalue density corresponding to $V_m$ vanishes
at the critical point $\lambda=0$ as $\lambda^{2m}$. To obtain the
most general scaling behaviour, we look for the potentials resulting
in $\rho(\lambda) \sim \lambda^{2m+1}$ for $\lambda \rightarrow 0$.
Such $\rho(\lambda)$ are not positive definite and we will introduce
them through perturbations around some $V_m (\lambda)$,
 in the spirit
of \ref\HNII{H. Neuberger, ``Regularized String and Flow equations,''
Rutgers preprint RU-90-50.}. The required behaviour is
\eqn\pertden{\eqalign{
\tilde \rho_n(\lambda) &= 0 \quad {\rm for} \quad |\lambda|>1,\cr
\tilde \rho_n(\lambda) &\propto \lambda^{n-1} \quad {\rm for} \quad
\lambda\rightarrow 0,\cr
&\propto (1-\lambda)^{1/2} \quad {\rm for} \quad \lambda\rightarrow
1^-,\cr
&\propto (1+\lambda)^{1/2} \quad {\rm for} \quad
\lambda\rightarrow -1^+,\cr
&\int_{-1}^1 d\lambda \tilde \rho_n(\lambda) =0.\cr}}
This is satisfied by
$$\tilde \rho_n(\lambda) = {d\over
d\lambda}[\lambda^n(1-\lambda^2)^{3/2}],\quad n=1,2,\ldots$$
and the corresponding potential is
$$\tilde V_n = v(n) N^{c(n)} \lambda^n (\lambda^2-1)^{3/2}|_+,$$
with $v(n)$ and $c(n)$ constants to be fixed.

The scaling ansatz appropriate to the potential $V_m$ is
\eqn\ansatz{\eqalign{R_n &= r_c + (-1)^n a^{1/m} f(x) + \cdots,\cr
S_n &= s_c + (-1)^n a^{1/m} i g(x) +\cdots,\cr
{n\over N} &= 1-a^2 (x-\mu), \qquad Na^{2+1/m} = 1.\cr}}
$s_c$ is of the order of odd perturbations, which are supressed by a
power of $N$. The orthonormal polynomials $\CP_n
(\lambda)$ tend to two scaling functions, depending on parity:
\eqn\polys{\eqalign{(-1)^n\CP_{2n}(a^{1/m}\lambda) &= a^{1/2m}
p_+(x,\lambda),\cr
(-1)^n\CP_{2n+1}(a^{1/m}\lambda) &= a^{1/2m}
p_-(x-a^{1/m},\lambda).\cr}}
{}From \ansatz\ and \polys\ it follows that the multiplication by the
eigenvalue $\lambda \equiv a^{1/m} \tilde \lambda$ is represented on
$\Psi = \left(\matrix{p_+\cr p_-\cr}\right)$ by a $2\times2$ matrix
(after a rescaling of $f$ and $g$ and a rotation of $\Psi$):
\eqn\laxa{\tilde \lambda \CJ_3 \Psi = (D + q)\Psi,}
where $D\equiv d/dx$, $q=f(x) \CJ_1 + g(x) i\CJ_2$, and
$\CJ_i$ are matrices satisfying $[\CJ_i,
\CJ_j] = i\varepsilon_{ijk} \CJ_k$. Finally, redefining $\Psi
\rightarrow \CJ_3 \Psi$ we obtain (dropping tilde from $\tilde
\lambda$)
$$\lambda \Psi = 4(D+q)\CJ_3\Psi \equiv A\Psi.$$
We will need the resolvent
$$R\equiv {1\over (D+q)\CJ_3 - {\lambda\over4}}.$$
In their work on the resolvents of matrix differential operators
\ref\GDmatrix{I.M. Gelfand and L.A. Dikii, ``The Resolvent and
Hamiltonian Systems,'' Funct. Anal. and Appl.}\ Gelfand and Dikii show
that $R$ satisfies
\eqn\reseq{\CJ_3 R' = [\CJ_3 R, q-\lambda\CJ_3],}
and that it has an asymptotic expansion
$$R(x,\lambda) = \sum_{k=0}^\infty R_k \lambda^{-k}.$$
{}From \reseq\ it is clear that, up to constants, the most general
expansion of $R$ is
\eqn\resexp{\CJ_3 R = \sum_{k=0}^\infty \grad\Ht_{k-1}
\lambda^{-k},}
where $\grad\Ht_k \equiv -\CJ_1 G_k -i\CJ_2 F_k + \CJ_3 H_k$.
Plugging \resexp\ into \reseq\ results in the recursion relations
determining $G_k$, $F_k$, $H_k$. Setting $G_{-1} = F_{-1} =0$, $H_{-1}
=1$, we obtain
\eqn\recura{ \eqalign{
        F_{k+1} &= G'_k + g H_k \cr
        G_{k+1} &= F'_k + f H_k \cr
        H'_k &= g G_k - f F_k,}}
for all $k\ge 0$. The first few are
\eqn\exampa{ \eqalign{
F_0 &=g,\cr
F_1 &=f',\cr
F_2 &=g''+{1\over 2}g(g^2-f^2),\cr}\quad \eqalign{G_0 &=f, \cr
G_1 &=g',\cr
G_2 &=f''+{1\over 2}f(g^2-f^2),\cr}\quad \eqalign{H_0 &=0, \cr
H_1 &={1\over 2}(g^2-f^2) ,\cr
H_2 &=g f'-f g', \cr}}
$$\eqalign{F_3 &=f'''+{3\over 2}f'(g^2-f^2),\quad
G_3=g'''+{3\over 2}g'(g^2-f^2), \cr
H_3 &=g g'' - {1\over 2}(g')^2 - f f'' +{1\over 2}(f')^2 +
{3\over 8}(g^2-f^2)^2 .\cr}$$
This determines the expansion of the resolvent
$$R=\sum_{k=0}^\infty (-i\CJ_2 G_{k-1} - \CJ_1 F_{k-1} +
H_{k-1}/4) \lambda^{-k}.$$
There is an $SO(1,1)$ symmetry in our system coming from possible
choices in the original definition of $q$. Observables will be
invariant under this symmetry.

A related derivation of \recura\ would start from two Hamiltonian
structures defined in
\ref\DrS{Drinfeld and Sokolov, ``Equations of Korteweg-de Vries
type and simple Lie algebras,'' Sov. Jour. Math. (1985)1975}
for the system with the Lax operator
\eqn\defl{L=D + q -\lambda \CJ_3.} For the flow in the coupling $t_k$,
with dimension $k$, their compatibility gives
\eqn\flows{{d\over dt_k}q = [\grad \Ht_k,\CJ_3] =
[\grad \Ht_{k-1},D+q],}
which is easily seen to result in \recura.

We are now ready to start determining the observables. We will
calculate the 2-point functions of the puncture operator $P$ with the
scaling operators $\sigma_n$ corresponding to the perturbations by
$\tilde V_n$. We will follow the presentation of \HNII, beginning with
the one-point function
\eqn\onepoint{{\partial F\over\partial t_n} = v(n) N^{c(n)} \oint_\CC
{dy\over 2\pi i} y^n (y^2-1)^{3/2}|_+ tr\biggl({1\over y-A}
\Pi_N\biggr),}
where
$$(\Pi_N)_{ij}=\cases{\delta_{ij} &if $0\le i,j\le N-1$,\cr
0& otherwise,\cr}$$
and $\CC$ surrounds all of the eigenvalues of $A$. By deforming $\CC$
to a circle at infinity one can see that the subscript $+$ in
\onepoint\
can be dropped. To obtain a two-point function, we study a variation
$\delta \partial F/\partial t_n$ and use \HNII:
$$\eqalign{\delta tr\biggl({1\over y-A} \Pi_N\biggr)
& = {1\over2}tr\biggl({1\over y-A} [S(\delta V(A)), \Pi_N]\biggr),\cr
S(B)_{ij} &= \varepsilon (i-j) B_{ij},\cr
\varepsilon (j) &= \hbox{sign of}\ j,\quad j\not= 0,\cr
&=0,\quad j=0.\cr}$$
We choose the simplest even operator as the puncture operator. The
corresponding variation of the potential is $\delta V(A) = N^{c_P}
A^2$, and we obtain (to leading order in $N$):
\eqn\twopoint{\eqalign{{\partial^2 F\over \partial x\partial t_n} &=
{v(n) \over2} N^{c_P+c(n)} \oint_\CC
{dy\over 2\pi i} y^n (y^2-1)^{3/2} tr\biggl({1\over y-A}
[S(A^2), \Pi_N]\biggr),\cr
&=v(n) r_c N^{c_P+c(n)} \oint_\CC
{dy\over 2\pi i} y^n (y^2-1)^{3/2} \biggl[\biggl({1\over
y-A}\biggr)_{N,N-2} + \biggl({1\over
y-A}\biggr)_{N-1,N+1}\biggr].\cr}}
The scaled quantities are
\eqn\scavar{\eqalign{y&=a^{1/m} \lambda ,\cr
A &= a^{1/m} 4(D+q)\CJ_3,\cr
|2n\rangle &= (-1)^n a^{1/2m} |x,+\rangle,\cr
|2n+1\rangle &= (-1)^n a^{1/2m} |x-a^{1/m},-\rangle,\cr}}
and \twopoint\ becomes (again to leading order)
\eqn\observ{\eqalign{
{\partial^2 F\over \partial x\partial t_n} &=
v(n) {r_c\over 4} a^{(n+1)/m} N^{c_P+c(n)} \oint_\CC {d\lambda\over
2\pi i} \lambda^n \bigl(\langle x,+|R|x,+\rangle +
\langle x,-|R|x,-\rangle \bigr)_{x=\mu},\cr
&= v(n) {r_c\over 2} a^{(n+1)/m} N^{c_P+c(n)} H_n(\mu).\cr}}

The simplest non-trivial case in this formalism is $m=1$ potential
\mceven\ with simplest odd perturbation \nappi
\eqn\simplepot{V = V_1 + Nae tr(\phi).}
(Note that, in the context of the unitary matrix model as
2D lattice Yang-Mills theory this coupling corresponds to
a discretized version of ${\rm log}\,{\rm det}U\,{\rm mod}2\pi$, and
hence corresponds to a lattice version of the theta angle
\ref\seiblatt{N. Seiberg, ``Topology in Strong Coupling,''
Phys. Rev. Lett. {\bf 53}(1984)637.}.)
Using \ansatz\ with $s_c = -ae/16$ in the recursion relations gives
string equations\foot{To obtain these string equations we rescale $e$,
$f$ and $g$ and shift $x$. In order to compare with \exampa\ the
rescalings of $f$ and $g$ have to be the same as the ones used to
obtain \laxa.}
\eqn\astring{\eqalign{0&= xf + eg' + 2(f''+{1\over2} f(g^2-f^2)),\cr
0&= xg + ef' + 2(g'' + {1\over2} g(g^2-f^2)).\cr}}
The specific heat is
\eqn\spechti{
\vev{PP} = -{x\over 4} + {e^2\over 48} + {1\over4}(g^2-f^2).}
The first two terms are non-universal and the
universal piece is given by $H_1/2$. Therefore, we can identify
$x\equiv t_1$, $\sigma_1\equiv P$. This identification shows that, in
order to obtain a finite result in \observ, we need
$c(n)=n/(2m+1)$.\ \foot{The factor of $Na$ in front of $tr(\phi)$ in
\simplepot\ is
precisely $N^{2/3}$ appearing in $\tilde V_2$. We use the fact that
the scaling limit of a {\it generic\/} odd (even) potential like
$tr(\phi)$ ($tr(\phi^2)$) is given by $\tilde V_2$ ($\tilde V_1$).}
Finally, for later convenience, we choose $v(n)=1/r_c$ for every $n$.
In conclusion, in the scaling limit of the two-cut matrix model we
have obtained the following observables
\eqn\inflow{
{\partial^2 F\over \partial x\partial t_n} \equiv \vev{P\sigma_n} =
\half H_n.}

The odd flows $F_{2k}$ and $G_{2k}$ have terms with no derivatives.
However, an insertion of an odd operator is still suppressed by
$1/N$.

To derive the string equations we will use
\eqn\seqrec{n=R_n^{1/2} \int V' \CP_n \CP_{n-1} = R_n^{1/2}\oint_\CC
{dy\over 2\pi i} V'(y) \biggl( {1\over y-A}\biggr)_{n,n-1},}
\eqn\fullpot{V'=Nk(m)y^{2m} (y^2-1)^{1/2}|_+ +
\sum_{k=1}^{2m} t_k{1\over r_c} N^{{k\over 2m+1}}y^{k-1}
(y^2-1)^{3/2}|_+ +\cdots,}
In \fullpot\ $+\cdots$ denotes terms of higher order in $a$ upon
introduction of scaling variables through $n = N(1-a^2 x)$ and
\scavar.

We want to replace the $y$-integral along $\CC$, a contour surrounding
all of the eigenvalues of $A$, by an integral over $\lambda$, i.e. over
infinitesimally small $y$'s. An obstacle to passing to a
$\lambda$-integral comes from the $A$-eigenvalues outside of the
critical region. Their contributions will not have scaling dependence
on coupling constants, but can be large ($\CO(N)$) nonetheless.

In order to be able to neglect consistently
eigenvalues of $A$ at large
$y$, instead of scaling \seqrec, we
will scale its derivative with respect to $t_l$, $l\not=1$. Namely,
the $t_l$ dependence of $V$ appears only for $y$ very small
($\CO(a^{1/m})$), while for $y\sim \CO(1)$ it is suppressed by a power
of (large) $N$. The result is
\eqn\derseq{0={\delta\over \delta t_l}\oint {d\lambda\over 2\pi i}
\sum_k k t_k \lambda^{k-1} (\mp)\langle x,\pm|\sum_{r=0}^\infty
(-i\CJ_2 G_{r-1} -\CJ_1 F_{r-1} +H_{r-1}/4) \lambda^{-r}
|x,\mp\rangle,}
where the upper (lower) sign corresponds to $n$ even (odd). After
doing the integral and taking the $1,2$ ($2,1$) element of the
resulting matrix, we have
\eqn\resder{\eqalign{0 &=\sum_{k\ge 1} k t_k G_{k-1} \equiv S_g,\cr
0 &=\sum_{k\ge 1} k t_k F_{k-1} \equiv S_f.\cr}}
The integration constant, which is
independent of all $t_l$, $l\not= 1$, can be seen to be zero by
comparison with \astring. (It was shown in
\ref\minahan{J.A. Minahan, ``Matrix Models with Boundary Terms and
the Generalized Painlev\'e II Equation,'' UVA preprint;
``Schwinger-Dyson Equations for Unitary Matrix Models with
Boundaries,'' UVA-HET-91-04}\
that by modifying the action to include boundaries, or
``quark'' degrees of freedom one can introduce a nonzero
constant on the left hand side of \resder.)

The string equation generalizes \astring. Moreover, it may
be equivalently written as a flatness condition, generalizing that
in \CCGM:
\eqn\flat{\left[L,{d\over d\lambda}-M\right]=0}
where
\eqn\mmk{\eqalign{M &= \sum_{k\ge 1} k t_k M_{k-1},\cr
M_k &= \grad \hat H_{k-1} + \lambda\; \grad \hat H_{k-2} + \cdots +
\lambda^{k-1} \grad \hat H_0 + \lambda^k \CJ_3,\cr}}
and $M_0 = \CJ_3$. $M$
is chosen to make the left hand side of \flat\ independent of $\lambda$,
as described in \DrS.

The flatness condition \flat\ can be interpreted as the isomonodromic
deformation condition for the monodromy defined by a solution of
\eqn\meq{\biggl( {d\over d\lambda} - M\biggr) \Psi =0.} Furthermore,
the partition function is again the isomonodromic $\tau$-function. The
proof of these statements follows the steps already outlined in
\nref\mr{G. Moore, ``Geometry of the string equations,''
Commun. Math. Phys. {\bf 133}(1990)261; ``Matrix Models of
2D Gravity and Isomonodromic Deformation,'' in
{\it Common Trends in Mathematics
and Quantum Field Theories} Prog. Theor. Phys. Suppl. {\bf 102}
(1990)255.}\
\refs{\mr,\CCGM}.
\foot{The form of recursion relations useful in the
proof is \recura\ with the
third equation replaced by
$$H_{k+1} = {1\over2} \sum_{r=0}^k (F_r F_{k-r} - G_r G_{k-r} - H_r
H_{k-r}).$$} Moreover, the solutions to the string equations
chosen by the matrix model will in general lead to nontrivial
Stokes data. As described in \mr\ this implies that the $\tau$
function does not correspond to a point in the Segal-Wilson
Grassmannian, although it does correspond to a point in the
Sato Grassmannian.

\newsec{Loop Equations.}

The solutions of one-matrix models satisfy Schwinger-Dyson
equations which can be derived by variation of the functional integral
\ref\WM{S.R. Wadia, ``Dyson-Schwinger equations approach to the large-$N$
limit - model systems and string representation of Yang-Mills
theory,'' Phys. Rev. D {\bf 24} (1981) 970;
A. A. Migdal, Phys. Rep. 102 (1983) 199.}.
These have a double scaling limit
\nref\davl{F. David, Mod. Phys. Lett. {\bf A5} (1990) 1019.}\
\nref\FKN{M. Fukuma, H. Kawai and R. Nakayama, ``Continuum
Schwinger-Dyson equations and universal structures in two-dimensional
quantum gravity,'' Tokyo preprint UT-562, May 1990.}\
\refs{\davl,\FKN}\
which in the KdV-type systems can be rewritten as Virasoro constraints
on the partition function \FKN
\ref\DVV{R. Dijkgraaf, E. Verlinde and H. Verlinde, ``Loop equations
and Virasoro constraints in non-perturbative 2-D quantum gravity,'' Nucl.
Phys. {\bf B348} (1991) 435.}.

Since the {\it same} Schwinger-Dyson equations apply (before taking the
continuum limit) to the two-cut Hermitian matrix model,
and since the unitary matrix model has analogous Schwinger-Dyson
equations,
one would expect continuum loop equations to exist in this case as well.

Previously we had derived a subset of these loop equations (with two
non-zero couplings)  by directly scaling the matrix model Schwinger-Dyson
equations
\ref\CDM{ \v C. Crnkovi\'c, M. Douglas and G. Moore, ``Physical Solutions
for Unitary Matrix Models,'' Nucl. Phys. {\bf B360}(1991)507}.
Here we derive the complete loop equations from the string equation/mKdV
formalism.
The argument is similar to that used in \DVV\ for the KdV theories.
Using the flow equations,
the string equation can be reformulated as a differential constraint
on the partition function as a function of the couplings.
In principle we can then derive a series of constraints as follows:
the same recursion relation which takes the $n$'th flow to the $n+1$'st
can be applied to the $n$'th constraint to find a new constraint.

Since the constraint operators all annihilate the partition function,
they form a left ideal.  Therefore commutators of constraints are also
constraints.  Guided by our expectation that the constraints form a
Virasoro algebra, we will find a minimal set of constraints which generate
this algebra.  The completeness of the resulting algebra will then follow
if we can argue that there is a unique solution of the resulting
constraints.  This will be true in a certain sense to be discussed below.

The scaling operator corresponding to the coupling $t_n$ will be called
$\sigma_n$.  $\sigma_1$ will also be called $P$, though one should
probably
think of the $P$ as standing for ``primary" and not ``puncture" for
reasons explained below.
The string equation is now $S_f=S_g=0$, and our constraints will be
derived from combinations of these.

We start with
\eqn\lminusone{ \eqalign{
        0 &= g S_g - f S_f \cr
        &= \sum_k k t_k \vev{\sigma_{k-1} \P \P} \cr
        &=D^2  \sum_{k\ge 2} k t_k \vev{\sigma_{k-1}} ,}}
which we interpret as the second derivative of an $L_{-1}$ constraint.
The explicit dependence on $t_1$ has cancelled out.

The recursion relations imply that $g G_{k+1}-f F_{k+1} = g F'_k-f G'_k$,
so we can get another equation from
\eqn\lzero{ \eqalign{
        0 &= g S'_f - f S'_g \cr
        &= \sum_k k t_k \vev{\sigma_{k} \P \P} + (g^2 - f^2) \cr
        &= \sum_k k t_k \vev{\sigma_{k} \P \P} + 2 \vev{\P \P} \cr
        &=D^2  \sum_k k t_k \vev{\sigma_{k}} }}
the second derivative of an $L_0$ constraint.

We can continue in this vein using the recursions to construct
$t_k d/dt_{k+n}$ in terms of differential polynomials of the string
equations.
We will skip $L_1$ (which is similar) and proceed to
the $L_2$ constraint, since it will generate the complete
algebra.  Using the recursions three times, we have
\eqn\itwo{ \eqalign{
        I_2 &= \sum_k k t_k \vev{\sigma_{k+2} \P \P} \cr
        &={1\over2}  \sum_k k t_k (g G_{k+2} - f F_{k+2}) \cr
        &={1\over2}    \sum_k k t_k (g F'_{k+1} - f G'_{k+1}) \cr
        &={1\over2}    \sum_k k t_k (g G''_k - f F''_k + (g^2-f^2)H'_k
                     + (gg'-ff')H_k). \cr
        &={1\over2}  \sum_k k t_k (g F'''_{k-1} - f G'''_{k-1} +
        X_{k-1} + (g^2-f^2)H'_k + (gg'-ff')H_k). \cr}}
with $X_{k-1} = g (f H_{k-1})''-f (g H_{k-1})''$.
The $X$ piece is identically zero by the $L_{-1}$ constraint \lminusone\
and its derivatives:
\eqn\dlmone{ D^n \sum_{k\ge 2} k t_k \vev{\sigma_{k-1}} =
         \sum_k k t_k D^{n-1} H_{k-1}.}

Using $S'''_f=\sum k t_k F'''_{k-1} + 3 g''=0$ (resp. for $g$), this is
\eqn\itwob{
        I_2 = -{3\over2}(gg''-ff'') + 2\sum_k k t_k (2\vev{\P\P}
\vev{\sigma_k \P\P}+  \vev{\P\P\P}\vev{\sigma_k \P}) }
and using derivatives of the $L_0$ constraint
\eqn\itwoc{
        I_2 = -8\vev{\P\P}^2 - 2\vev{\P\P\P}\vev{\P} -{3\over2}
 (gg''-ff'').}
The other terms we expect in an $L_2$ constraint are
\eqn\pfour{ \vev{\P\P\P\P} = {1\over2}\bigl(gg''-ff'' + (g')^2 -
(f')^2\bigr), }
\eqn\sthree{ \vev{\sigma_3 \P} = {1\over2}\bigl(g g''-ff''-{1\over 2}
    (g')^2+{ 1\over 2}(f')^2 + {3\over 8}(g^2-f^2)^2\bigr) .}
These combine nicely to give
\eqn\ddltwo{
0=\sum_k k t_k \vev{\sigma_{k+2}\P\P} + 2\vev{\sigma_3 \P}+
\vev{\P\P\P\P}+2\vev{\P\P}^2+
       2 \vev{\P\P\P}\vev{\P}. }
Defining $Z=e^F$, this is the second derivative of
the expected constraint $L_2 Z=0$ with
\eqn\ltwo{ L_2 = \sum_{k \ge 1} k t_k {\partial\over\partial t_{k+2}}
                + {\partial^2\over\partial t_1^2}. }

Note that to get to \itwob\ we assumed that the undifferentiated
$L_{-1}$ and $L_0$ constraints were satisfied, i.e. that they
contained no integration  constant. Under this assumption, we can now
write the result as Virasoro constraints for a two-dimensional free
massless boson.  Define
\eqn\boson{\partial\phi(z)\equiv \sum_{k\ge 1}
k z^{k-1} t_k +  \sum_{k\ge 1} z^{-1-k}
{\partial\over\partial t_k} ,}
an untwisted boson; then the
constraints are the modes $L_n$ for $n\ge -1$ of the usual untwisted
stress-tensor $(\partial\phi)^2$ with zero ground-state energy.

The integration constants in the constraints must be compatible with
the Virasoro algebra, otherwise the commutator of two constraints
would generate new constraints which would force $Z$ to be trivial.
An obvious possibility would be to include the zero mode of the boson
as well,
\eqn\zboson{\partial\phi(z)\equiv \sum_{k\ge 1} k z^{k-1} t_k + z^{-1} q +
                \sum_{k\ge 1} z^{-1-k} {\partial\over\partial t_k} .}
$q$ would be a new parameter of the theory, not associated with any
operators or flows.
In the ``usual" phase of the theory with all odd couplings set to zero,
it is clear that $q=0$, since all the other terms in the $L_{-1}$
constraint are zero, but we will find a use for non-zero values shortly.

It should also be possible to derive these loop equations directly
by double scaling the finite $N$ Schwinger-Dyson equations.
In particular, the theory of the boson \zboson\
should emerge from the formulation of the
lattice theory using second quantized fermions.
Unfortunately, the currently available derivations of this
``field theory on the spectral curve,'' are merely heuristic.

\newsec{A Topological Phase}

In the series of multicritical models including pure gravity,
described by the string equation
\eqn\kstring{ x = t_1 u + t_2 ( u''+3u^2) + \ldots ,}
the simplest point is not pure gravity but rather the ``pure topological"
theory with $t_1=1$ and $t_n=0$ for $n>1$.
This theory has a conserved charge
(``ghost number") with a distinct background charge
for each genus surface, so a specific correlation function can be non-zero
for at most one genus.
In particular $\vev{\sigma_0 \sigma_0 \sigma_0}=u'=1$ is non-zero only on
the sphere.

With purely even couplings, there is no analogous model in the mKdV
hierarchy, but with odd couplings there is a precisely analogous model
with $t_2=1$ and $t_n=0$ for $n>2$.  The string equation is then
\eqn\tstring{ \eqalign{ x g + f' &= 0 \cr x f + g' &= 0 .}}
with solution
\eqn\tsol{
\eqalign{ g + f &= C_1 e^{-x^2/2} \cr g - f &= C_2 e^{x^2/2} .}}
$C_1$ and $C_2$ are constants of integration.
Although these solutions are trancendental, physical quantities are not:
for example,
\eqn\toneone{ \vev{\sigma_1 \sigma_1} = {1\over 2}(g^2-f^2) =
{1\over 2}C_1 C_2 ,}
so this model appears to have ``$\gamma_{string}=0$".
(The Baker functions, which are Airy functions in topological gravity,
are parabolic cylinder functions in this theory.)

The $SO(1,1)$ symmetry insures that all correlation functions in this
model are polynomial in $x$.
Assign the combinations $f+g$ and $f-g$
$SO(1,1)$ charge $1$ and $-1$; then correlation functions must be neutral.
All charge-neutral polynomials in these functions are constant, while
appearances of $d/dx$ will produce positive powers of $x$.

The structure of the topological phase is seen  most clearly in the
loop equations.  Just as for topological gravity, expanding the loop
equations around these couplings gives a series of recursion relations,
one for each operator in the theory, expressing an operator insertion
in terms of correlators of lower total degree.
For example, the $L_{-1}$ constraint gives us
\eqn\tlmone{ \vev{\sigma_1 \prod_i \sigma_{n_i}} =
        - {1\over 2}\sum_i n_i \vev{\sigma_{n_i-1}\prod_{j\ne i}
                \sigma_{n_j}} }
(where $\sigma_0\equiv 0$) for all cases except one:
\eqn\tlmonedeg{ \vev{\sigma_1\sigma_1} = -{q\over 2}.}
To get non-trivial answers in the topological phase, we must take the
parameter $q=-C_1 C_2$ non-zero.

Just as for the KdV models, the existence of a phase in which all
operator insertions satisfy recursion relations, means that the
expansion of the partition function to all orders is uniquely determined
by the constraints.  They can therefore have only one analytic solution.
Unlike the KdV models, since the parameter $q$ is not a coupling of
the model, we cannot flow from this topological phase to the
phases with $q=0$.
This is probably connected with the difficulty of defining this pure
topological phase in the matrix model -- our odd perturbations of the
potential \pertden\ really only make sense as perturbations of
even potentials.
On the other hand the structure of the pure topological phase seems very
clear from the solution, so we will not let this stop us from studying it.

The ghost number conservation law follows most directly from the
loop equations.  If we write them in terms of our boson $\partial\phi(z)$,
they are homogeneous in $z$, so the ghost number assignment of each
operator
and coupling must be just the associated power of $z$ in a mode expansion,
up to a possible constant shift.  We can use this constant shift to make
the ghost number of one coupling of our choice zero.  If we choose $t_2$,
setting $t_2=1$ will not break ghost number conservation.

For topological gravity this is the end of the story.  The special role
of $t_2$ in this theory is that it is the only choice of non-zero coupling
for which all operator insertions have an associated recursion relation.
In the present case, both $t_2$ and $q$
must be set non-zero, and they have different ghost number, $q$ having
that of ``$t_0$".
A simple example of a correlation function which shows that $q$
must be assigned ghost number is
\eqn\corex{ \vev{\sigma_3 \sigma_4} =
-48 \lambda^5 q x - 144 \lambda q^3 x + 168 \lambda^3 q^2 x^3
- 24 \lambda^5 q x^5}
where $\lambda$ counts derivatives (powers of $1/N$).

We can maintain a ghost number conservation law if we take $q$ non-zero
but treat it as the coupling of a new operator in the theory.
If the operator $\sigma_n$ is given ghost number $2-n$, then $q$
will couple to an operator $Q$ with ghost number $2$.
Now the basic correlation function on the sphere is
\eqn\bascor{\vev{\sigma_1 \sigma_1 Q} = -{1\over 2}.}

$Q$ is an unusual operator in that it does not have descendants.
The ghost number counting would suggest that it could be renamed
$\sigma_0$ and that the other operators are its descendants,
but this is misleading.
It clearly appears in a very different way in the string equation;
in the loop equations, it is different in that its conjugate
$d/dq$ does not appear.
The necessity to include this operator is the most serious difference with
the topological gravity coupled to matter of \witten.

\newsec{Macroscopic Loops for Unitary-Matrix Models}

It has recently been shown that the study of macroscopic
loop amplitudes
\ref\bdss{T. Banks, M. Douglas, N. Seiberg, and S. Shenker,
``Microscopic and macroscopic loops in non-perturbative two
dimensional gravity,''
Phys. Lett. {\bf B241}(1990)279}
is particularly well-suited for the
comparison of continuum-Liouville and matrix model
formulations of 2D quantum gravity
\ref\mss{G. Moore, N. Seiberg, and M. Staudacher, ``From
Loops to States in 2D Quantum Gravity,'' Rutgers/Yale
preprint RU-91-11/YCTP-P11-91}
\ref\related{G. Moore and N. Seiberg,
``From Loops to Fields in 2D Quantum Gravity,'' Rutgers/Yale
preprint RU-91-29/YCTP-P19-91}.
In particular,
the Wheeler-DeWitt equation plays a central role in
making such comparisons. In this appendix we
derive analogous macroscopic loop amplitudes for
the unitary-matrix model, and show that they too
obey a Wheeler-DeWitt-like equation. This equation
gives a hint about the proper worldsheet interpretation
of these theories.

The macroscopic loop operator is defined to be
$\Psi^\dagger e^{-\ell L}\Psi$ in the fermionic
formulation, where $\Psi$ is a two-component
fermi field (obtained from scaling the even and
odd indexed orthogonal polynomials) and $L$ is
the Lax operator representing multiplication by
$\lambda$. The one-point function is therefore
\eqn\onepti{\langle w(\ell)\rangle=\int_{t_1}^\infty
dx\langle x|{\rm tr}\Biggl\{e^{-\ell\pmatrix{D&-f-g\cr
f-g&-D\cr}}\Biggr\}|x\rangle
}
We obtain the genus zero approximation by evaluating
the matrix element with the WKB approximation to get
\eqn\oneptii{\eqalign{
\langle w(\ell)\rangle&=\int_{t_1}^\infty dx \int_{-\infty}^\infty
dp~{\rm tr}\Biggl\{e^{-\ell\pmatrix{ip& -f-g\cr f-g&-ip\cr}}\Biggr\}\cr
&=\int_{t_1}^\infty dx\int_{-\infty}^\infty dp
\biggl(e^{-\ell\sqrt{g^2-f^2-p^2}}+e^{\ell\sqrt{g^2-f^2-p^2}}\biggr)\cr
&=2\int_{t_1}^\infty dx\int_{-\infty}^\infty dp~ {\rm
cos}(\ell\sqrt{p^2-H_1})\cr
&=-\pi\int_{t_1}^\infty dx h_1^{1/2}J_1(\ell h_1^{1/2})\cr}
}
where $h_1\equiv -H_1$ is positive (for large $x$)
and $J_1$ is a Bessel function.
Using the flow equations, specialized to the case of
genus zero, we therefore obtain the wavefunctions of the
even flow operators:
\eqn\wvfun{\eqalign{
\langle \sigma_{2k-1}w(\ell)\rangle&=-\pi \ell\int_{t_1}^\infty dx
h_1^{k-1} {\p h_1\over \p x} J_0(\ell h_1^{1/2})\cr
&=-\pi \ell\int_{h_1}^\infty dy
y^{k-1} J_0(\ell y^{1/2})\cr}
}
The operators for the odd flows have zero wavefunction
at tree level.
The integral in \wvfun\ only converges in a distributional sense. We
may imagine regularizing the original macroscopic loop
operator so as to obtain an additional factor of $e^{-\epsilon \sqrt{y}}$
in \wvfun. It is then easy to show that \wvfun\ and subsequent
formulae below possess a finite unambiguous limit as
$\epsilon\rightarrow 0^+$. In particular, using the identities
${2\nu\over z} J_{\nu}(z)=J_{\nu+1}(z)+J_{\nu-1}(z)$ and
\eqn\identity{
\lim_{\epsilon\to 0^+}\int_1^\infty dy e^{-\epsilon \sqrt{y}}
(y-1)^{k-1}J_0(a\sqrt{y})=\Gamma(k) 2^k (-a)^{-k}J_{k}(a)
}
we may prove (as in \mss)
the existence of a linear, upper-triangular,
analytic change of operators $\sigma_\nu\to \hat\sigma_\nu$
which have the wavefunctions:
\eqn\dfsghat{
\psi_\nu(\ell)\equiv
\langle\hat\sigma_{\nu} w(\ell)\rangle = h_1^{\nu/2}J_{\nu}(\ell
h_1^{1/2})
}
for $\nu=2k-1$, with $k=1,2,...$.

The wavefunctions of the $\hat\sigma$
operators satisfy the Bessel equation, which, in
the one-cut phase has been interpreted as the Wheeler-DeWitt
equation for gravity coupled to conformal matter in the minisuperspace
approximation. Assuming a similar interpretation holds in
this case we see that the Bessel equation should be interpreted
as
\eqn\wdweq{
\Biggl[\bigl(\ell{\p\over\p\ell}\bigr)^2-{\mu\over\gamma^4}\ell^2-
{8\over\gamma^2}\bigl({Q^2\over 8}+\Delta-1\bigr)\Biggr]\psi_\nu=0
}
where $\gamma,Q,\mu$ are parameters in the Liouville theory,
following the conventions of
\ref\nati{N. Seiberg, ``Notes on Quantum Liouville Theory
and Quantum Gravity,'' in {\it Common Trends in Mathematics
and Quantum Field Theories} Prog. Theor. Phys. Suppl. {\bf 102}
(1990) 319.},
and $\Delta$ is the dimension of the matter field being
dressed. It follows that the cosmological constant must
be considered to be negative. Accordingly, for a convergent
path integral
the worldsheet should have a Minkowskian signature
\ref\natiunpub{The negative cosmological constant
theory in Minkowski space has been studied
in unpublished work of N. Seiberg. See also
E. Martinec and S. Shatashvili, ``Black
Hole Physics and Liouville Theory,'' Chicago preprint
EFI-91-22.}.
Moreover, it follows from \wdweq\ that the central
charge and a subset of dimensions of the matter theory
must satisfy
\eqn\condition{\eqalign{
\nu^2&={8\over \gamma^2}({1-c\over 24}+\Delta)\cr
\gamma^2&={1\over 6}(13-c-\sqrt{(1-c)(25-c)})\cr}
}
where $\nu$ is any odd integer.

We can now use \dfsghat\condition\ to learn about
the continuum theory. Let us assume that the structure
of wavefunctions \dfsghat\ is exactly analogous to that
described in \mss. In particular, we assume that the
multicritical phases of the matrix model
defined by $k=1,2,3,\dots$ are characterized by the identification
of the wavefunction of the cosmological constant with
$\psi_{\nu_0}$ for $\nu_0=2k-1$. Plugging into
\condition\ we find a series of matter theories
with central charges
\eqn\censer{c=1-{6\nu_0^2\over \nu_0+1}
}
Again using \condition\ we discover that the $\psi_\nu$
are wavefunctions of operators whose flatspace dimensions are
\eqn\polyweights{
\Delta_\nu={\nu^2-\nu_0^2\over 4(\nu_0+1)}
}

The above critical exponents are consistent
with the identification of the matter theory as a
special case of the $O(n)$ model
\ref\saleurpriv{We are grateful to H. Saleur for explaining
to us relevant results in the theory of dense
polymers and for many discussions related to the
unitary-matrix model.}. In particular, the
$O(n)$ model for $n=-2{\rm cos} \pi g$ is thought
\ref\sashaz{A. Zamolodchikov, unpublished.}
to have a hierarchy of multicritical points, for a fixed
$n$, with central charges
\eqn\oncenchr{c=1-{6(g-1)^2\over g}
}
Moreover, the spectrum of the model includes
operators with weights given by
\ref\saleur{
B. Duplantier and H. Saleur, Nucl. Phys. {\bf B290}(1987)291
}
\eqn\onweights{\Delta_\nu={\nu^2-(g-1)^2\over 4g}
}
for $\nu\in \IZ$. The operator algebra respects the
$\IZ_2$ grading defined by the parity of $\nu$.
It is possible to identify the $\IZ_2$ eigenspaces
as Ramond and Neveu-Schwarz sectors of a conformal
field theory
\ref\newsaleur{The fermionic description of dense
polymers and $O(n=-2)$ models, in particular the
detailed meaning of $R$ and $NS$ sectors will be
described in a forthcomming paper of H. Saleur.}.
In view of these results
\sashaz\saleur\newsaleur\ we may identify $g=\nu_0+1$, an
even integer, so that $n=-2$. Only the ``Ramond operators''
with $\nu$ odd have nonvanishing wavefunctions at
genus zero, but it is natural to identify all the $\sigma_\nu$ as
the gravitationally dressed versions of the $O(n=-2)$
operators with dimensions \onweights.

\newsec{No Conclusions}

In the previous sections we have described in detail the
solution of the unitary-matrix model, including
string equations, flows, Virasoro constraints and
macroscopic loop amplitudes. Returning to the
questions mentioned in the introduction, we can now make
some comparisons with candidate continuum formulations.
We will discuss three candidate theories below. Unfortunately,
in all three cases the available evidence
remains inconclusive.

The first candidate is 2D supergravity. The multicritical points
would correspond to supergravity coupled to the $(2,4k)$
superminimal matter theories, as already discussed in \CDM.
Evidence for this is (1) the structure of even and odd scaling operators
and their dimensions match the $(2,4k)$ series of super-minimal matter;
(2) correlators of even (Neveu-Schwarz) operators agree with those of
bosonic gravity at tree level
\ref\difdiskut{P. Di Francesco, J. Distler, and D. Kutasov,
``Superdiscrete series coupled to 2D supergravity,'' Princeton preprint
PUPT-1189, June 1990.};
(3) there is no space-time supersymmetry or Grassmann couplings,
as one might expect for supergravity coupled to low $c$ minimal models
with no possibility of GSO projection.

Clearly the above is not very compelling evidence.
One would like to have, for example,
tree level Ramond correlators or an understanding of the super-WdW
equation to make a more definite statement. Higher genus correlators could
also settle the question but problems
associated with sums over spin structures
make these difficult to calculate. In particular,
the calculation of the one-loop partition function from
the continuum theory could be revealing.
The contribution for the even spin structures has been carefully
determined in
\ref\berkleb{M. Bershadsky and I. Klebanov, ``Partition functions and
physical states in two-dimensional quantum gravity and
supergravity,'' HUTP-91/A002,PUPT-1236}.
There is a further point which is rather difficult to reconcile with
this interpretation.  The two-cut model in one dimension is the {\it same}
to all orders in perturbation theory as the one-cut model
\ref\kazrev{V. A. Kazakov, ``Bosonic strings and string field theories in
one dimensional targer space,'' to appear in the proceedings of the Carg\' ese
workshop on random surfaces and quantum gravity.}
\ref\mooredbl{
G. Moore, ``Double-Scaled Field Theory at c=1,'' YCTP-P8-91, RU-91-12}.
This would be rather surprising for $\hat c=1$ matter coupled to supergravity.
Nevertheless, the supergravity interpretation
has not yet been definitely ruled out.

The second candidate is dense polymers
or the $O(n=-2)$ fermionic model, (these two
being intimately related non-unitary matter theories
with $c=-2$ \saleur) coupled to Lorentzian metric gravity.
The principle evidence for this was described in
section five and follows from assuming that
the macroscopic loop amplitudes
satisfy the Wheeler-DeWitt equation of 2D gravity.
A further hint that this is the correct interpretation
comes from examining in detail the perturbation series
of the simplest unitary-matrix model using the
action in \Neuberger. The Feynman perturbation series
can be written as a sum over surfaces together with a sum
over self-avoiding loop configurations (the fermion loops)
on those surfaces. Unfortunately, the Boltzman weights for
the loops are similar to, but not precisely equal to
those of the $O(n)$
model and it is not clear if the difference is important at
criticality. Moreover,
in this interpretation it is also rather mysterious that the
partition functions at every genus should be positive \peri.
Finally, the $\IZ_2$-odd loops should be calculated and compared with
the $NS$ sector of the polymer problem.

The third candidate is a candidate for the topological
phase discussed in section four.
The similarity of the structure of the topological
phase to the KdV-type systems strongly suggests
that the theory has a topological interpretation.
Ghost number counting rules out the obvious possibility
that it is $Osp(2|1)$ topological gravity.
Another idea follows from recalling the identification of
$t_2$ with the ``Yang-Mills $\theta$ angle'' described
below \simplepot.
In the continuum this operator becomes simply
$\int {\rm tr}F$ which could serve as an action for
a topological Yang-Mills matter theory, which could
then be coupled to topological gravity
\ref\lpw{J.M.F. Labastida, M. Pernici, and E. Witten,
``Topological gravity in two dimensions,'' Nucl. Phys. {\bf B310}
(1988)611.}
\ref\monson{D. Montano and J. Sonnenschein, ``The topology
of moduli space and quantum field theory,'' Nucl. Phys. {\bf B324}
(1989)348.}\witten.
However, there are some troubles with {\it any}
candidate topological matter theory.
If we have a theory of
topological gravity coupled to topological matter,
then, as shown by
Dijkgraaf and Witten there must be a ``puncture
operator" $P$ which satisfies a ``puncture equation" expressing an
insertion of $P$ in terms of correlation functions of lower degree.
The natural candidate for the puncture equation
would be the $L_{-1}$ constraint.
However,
this interpretation would make all operators $\sigma_n$ descendants of
a single primary, $\sigma_1$, which seems incompatible with the
uniqueness of the known TFT with a single primary.
In addition, ghost number conservation required us to introduce a
new operator $Q$ with no descendants, which does not fit at all into
the topological gravity framework.

Although it is possible that a different interpretation of the model
exists, which is more compatible with the original topological gravity
framework, we did not find such a description and suspect that it does
not exist. If this is true, clearly it would be very interesting to
find the correct world-sheet topological description.

\bigskip

\centerline{\bf Acknowledgements}

It is a pleasure to thank T. Banks, V. Kazakov,
D. Kutasov, E. Martinec,
Y. Matsuo, H. Saleur, N. Seiberg,
S. Shenker, M. Staudacher, and A. Zamolodchikov
for several useful discussions
relevant to these matters. M.R.D. especially thanks the
Laboratoire de Physique Th{\'e}orique
de l'Ecole Normale Sup{\'e}rieure, where this work was
begun, for its warm hospitality.
G.M. also thanks the Rutgers Department of Physics for hospitality
where some of this work was done. \v C.C. and G.M. are supported
by DOE grant DE-AC02-76ER03075, and G.M. is supported by a
Presidential Young Investigator Award. M.R.D. is supported by
DOE grant DE-FG05-90ER40559, an NSF Presidential Young Investigator Award,
and a Sloan Fellowship.

\listrefs
\bye